# Evaluation of deep learning-based myocardial infarction quantification using Segment CMR software

Olivier Rukundo
Department of Clinical Physiology, Lund University,
Lund, Sweden
olivier.rukundo@med.lu.se

## ABSTRACT

This work evaluates deep learning-based myocardial infarction (MI) quantification using Segment cardiovascular magnetic resonance (CMR) software. Segment CMR software incorporates the *e*xpectation-maximization, *w*eighted intensity, *a* priori information (EWA) algorithm used to generate the infarct scar volume, infarct scar percentage, and microvascular obstruction percentage. Here, Segment CMR software segmentation algorithm is updated with semantic segmentation with U-net to achieve and evaluate fully automated or deep learning-based MI quantification. The direct observation of graphs and the number of infarcted and contoured myocardium are two options used to estimate the relationship between deep learning-based MI quantification and medical expert-based results.

## 1. INTRODUCTION

Segment CMR) is a complete software solution developed by MEDVISO in collaboration with Lund Cardiac MR group at Lund University, clinically approved for cardiac strain analysis [1], [2]. In this work, Segment software segmentation algorithm is updated to achieve and evaluate fully automated MI quantification or simply deep learning-based MI quantification. The MI reflects the cell death of cardiac myocytes caused by ischemia, which is the result of a perfusion imbalance between supply and demand [3]. Despite several algorithms so far developed for quantification purposes [4], there is still no clinical standard for quantification of the size of the MI [5]. In [5], authors concluded that the EWA algorithm might serve as a clinical standard for the quantification of MI in heart images imaged by LGE-CMR [5]. Also, authors stated that an automatic algorithm for MI quantification was implemented and incorporated in the freely available Segment software. To achieve deep learning-based MI quantification, the automated semantic segmentation of the myocardium is required. Note that there are works in which tracing the boundaries of objects of interest is still done manually. In some recent works (this included), the boundaries of objects of interest are automatically traced or segmented using deep learning means [6], [7], [8]. For example, in our previous work, the U-net was the choice of interest to achieve accurate segmentation of the myocardium [8]. U-net is the convolutional neural network architecture that outperformed the then best method sliding-window convolutional network and won many challenges [6]. In [8], the U-net proved to fit for accurate semantic segmentation of the myocardium, and previous experiments proved that it almost matched with manual ground truth segmentation results, with the smallest L2 regularization value. The rest of the paper is organized as follows: Part II briefly introduces and discusses the Segment CMR and EWA algorithm. Part III provides experimental results and discussions. Part IV gives the conclusion.

## 2. SEGMENT CMR SOFTWARE

MEDVISO AB products include Segment CMR and Segment Research software [1]. Segment CMR is a complete software solution for cardiac MRI analysis and a clinically approved software solution for cardiac strain analysis [1]. Segment Research is freely available for research purposes and it includes a broad range of analysis tools for MRI, CT, and myocardial perfusion SPECT images [1]. Both stand-alone and source code versions, developed in collaboration with Lund Cardiac MR group, are available at the MEDVISO AB website download page and Lund Cardiac MR group GitHub page [1], [9]. In [2], the Segment's creator shared preliminary information on the design and validation of the then freely available version of Segment. In this work, the 3.1 R8225 version of Segment was updated to evaluate deep learning-based MI quantification and it incorporated the EWA quantification algorithm. The EWA algorithm is an automatic algorithm for quantification of the size of MI imaged by late gadolinium enhancement (LGE)-magnetic resonance imaging (MRI). EWA algorithm is based on three major principles: Expectation-Maximization for intensity classification, the weighted summation of infarct size to account for partial volume effects according to pixel intensity, and a priori information utilized for pre-and post-processing. In [5], authors provided more details on the EWA algorithm and proved that it might serve as a clinical standard for the quantification of MI in heart images imaged by LGE-CMR.

## 3. EXPERIMENTAL RESULTS AND DISCUSSIONS

**A) Image datasets:** Here, the author used LGE-MRI datasets containing 3587 images of the size 128 x 128. Such images were exported from 24 image stacks using Segment CMR software (version 3.1.R8225), the design and validation of which was presented in [2]. Ground truth segmentation images were converted into masks with the help of the same Segment software. Each ground truth segmentation image or mask consisted of three classes corresponding to 255 (region-1), 128 (region-2), and 0 (region-3) labels. The original dataset was split into three datasets, namely training dataset (60%), validation dataset (20%), and testing dataset (20%).

**B) U-net settings and graphic card:** The U-net's training hyperparameters were manually adjusted based on the observation of the training graph with the possibility for new adjustments when 10% of all epochs were reached before the training validation accuracy reached 90% [8]. Here, U-net's training hyperparameters, manually adjusted, included the number of the epochs = 180, minimum batch size = 16, initial learning rate = 0.0001, L2 regularization = 0.000005. Adam was the optimizer. The execution environment was multi-GPU with Nvidia Titan RTX graphic cards. Data augmentation options used to increase the number of images in the dataset used to train the U-net were a random reflection in the left-right direction as well as the range of vertical and horizontal translations on the interval ranging from -10 to 10.

**C) Metrics and simulation software:** Given that deep learning-based or fully automated MI quantification requires automated segmentation or semantic segmentation with U-net, the quality of automated semantic segmentation using U-net against the manual ground truth segmentation is evaluated using class metrics such as classification accuracy, intersection over union (IoU), and mean (boundary F-1) BF scores. Here, the accuracy metric was used to estimate the percentage of correctly identified pixels for each class. The IoU metric was used to achieve statistical accuracy measurement that penalizes false positives and mean BF was used to see how well the predicted boundary of each class aligns with the true boundary or simply use a metric that tends to correlate with human qualitative assessment, respectively [10], [11]. The infarct scar volume and percentage, as well as the microvascular obstruction percentage scores are generated by Segment software version 3.1 R8225's EWA [3], [6] while the number of infarcted and contoured myocardium is directly counted from displayed images. Here, the simulation software was MATLAB R2020b, but Segment software works better with MATLAB R2019b.

**E) Evaluation of segmentation results:** With U-net, the achieved final training validation accuracy was 99.11%. The class metrics-based scores or semantic segmentation with U-net results are presented in the top-left corner of Figure 1. Note that each semantic segmentation with U-net image or output mask had also three regions. Now, based on all achieved scores presented, only the smallest or minimum score of semantic segmentation with U-net was slightly above 90%, which could be considered as a positive sign or good performance in this direction. Note that, the time taken to train LGE-MRI data was not the main concern, therefore relevant details were not included in this work.

**F) Evaluation of quantification results:** First, 24 image stacks were loaded in the Segment CMR software. Next, stacks of interest were selected consecutively. The plugin or updated Segment CMR segmentation algorithm was run to generate the fully automated or deep learning-based MI quantification results and compare them against medical experts-based MI quantification results. Figure 1-(top-right) shows the average infarct scar volume of all images in each one of the 24 stacks listed on the x-axis of figure 1. Based on medical experts-based MI quantification results, more than 50 % of the deep learning-based MI quantification results or simply network results were very close to the medical experts-based MI quantification results. Figure 1-(bottom-left) shows the average infarct scar percentage of all images in each one of the 24 stacks listed on the x-axis of figure 1. Here, more than 75 % of the network results were very close to the medical experts-based MI quantification results. Figure 1-(bottom-right) shows the average microvascular obstruction (MO) percentage of all images in each one of the 24 stacks, listed on the x-axis of figure 1. Here, more than 65 % of the network results were very close to the medical experts-based MI quantification results. Note that, here, the option of interest was to separate automated results by observation of the graphs to find fully automated results that were close to some extent to medical experts' results. Therefore, the numerical threshold values were not used because the direct observation was assumed to be enough to determine the closeness of scar (ml), scar (%), and mo (%), in both cases.

**G) Evaluation of the number of infarcted and contoured myocardium:** Figure 2 shows the example of one stack consisting of several cine images. Figure 2 (a) shows the medical experts-based results. Figure 2 (b) shows the deep learning-based results. Here, the endocardium boundary or contour is shown by the red outline. The epicardium contour is shown by the green outline. The infract contour is shown by the yellow outline. The microvascular obstruction is shown by the red outline inside the infarcted area. In Figure 2 (a) and Figure 2 (b), the numbers of images containing infarcted areas are equal. The same situation happened with the number of images having myocardial contours.

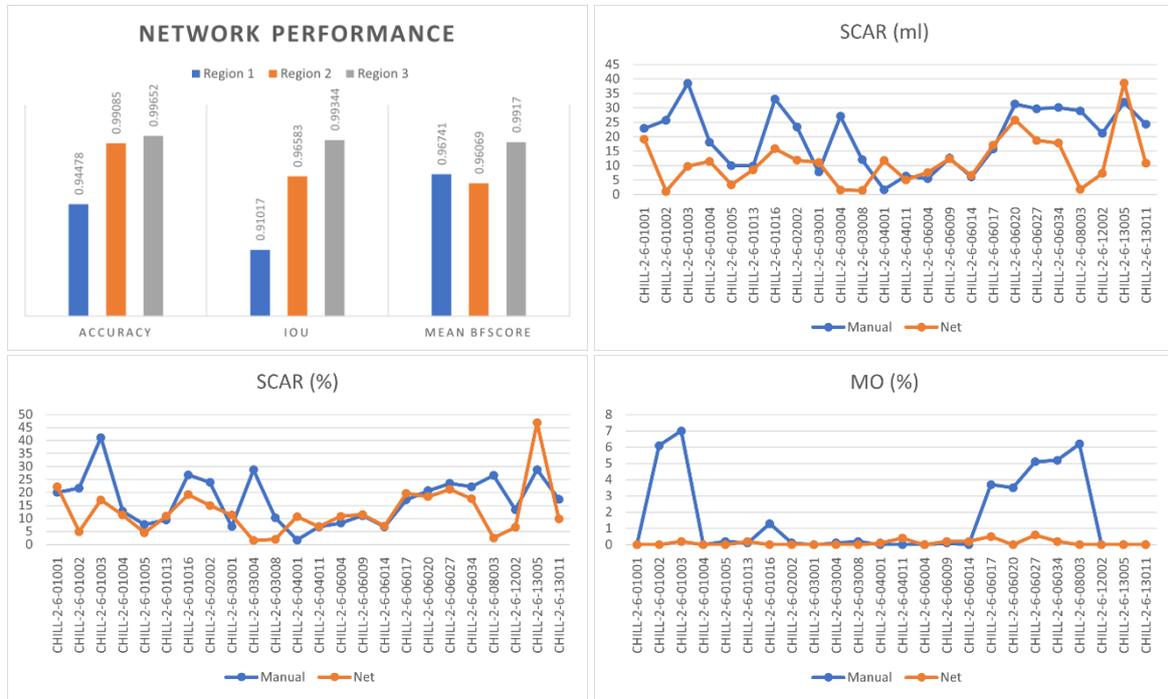

Figure 1: Performance of semantic segmentation with U-net and evaluation of deep learning-based MI quantification vs medical experts-based results

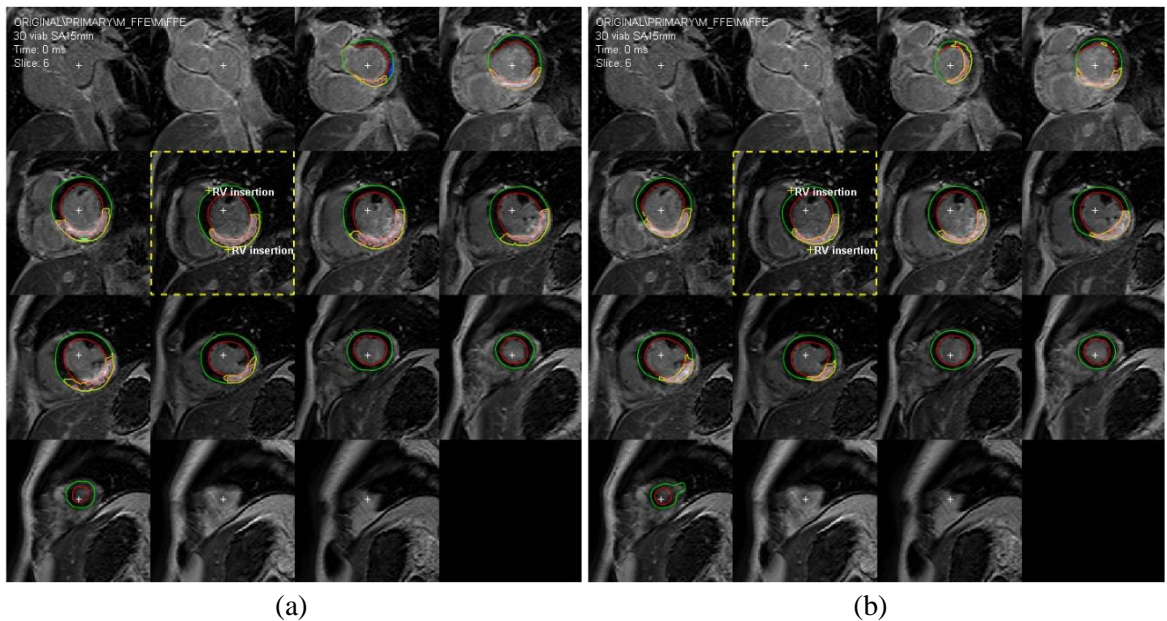

(a)                  (b)

Figure 2: Number of infarcted and contoured myocardium in (a) medical expert-based and (b) deep learning-based

## 4. CONCLUSION

In this evaluation, the final validation accuracy achieved (after training U-net) was 99.11% and the minimum semantic segmentation class metrics-based score was only slightly above 90%. As result, fully automated or deep learning-based MI quantification results showed that more than 50 % of the average infarct scar volume, 75% of infarct scar percentage, and 65 % of microvascular obstruction percentage were very close to the medical experts-based results. In addition, referring to a specific example (involving one stack) the deep learning-based method performed the same way as the medical experts-based method by achieving the same number of infarcted and contoured myocardium in images of interest displayed using the Segment CMR software. Future work may focus on exporting a very big training dataset as well as training data optimization to improve deep learning performances.